\documentclass[copyright,creativecommons]{eptcs}
\providecommand{\event}{CREST 2017 (invited talk)} % Name of the event you are submitting to
\usepackage{breakurl}             % Not needed if you use pdflatex only.
\usepackage{underscore}           % Only needed if you use pdflatex.

\usepackage[latin1]{inputenc}

\usepackage{graphicx}
\usepackage{epstopdf}
\usepackage[usenames]{color}
\usepackage{amsmath}
\usepackage{amsthm}
\usepackage{amsfonts}
\usepackage{amssymb}
\usepackage{makeidx}
\usepackage{stmaryrd}
\usepackage{framed}
\usepackage[ruled,vlined]{algorithm2e}
\usepackage{multirow}
\usepackage{tikz}
\usetikzlibrary{shapes.geometric,decorations.pathmorphing}
	\usetikzlibrary{automata,positioning,calc,through,backgrounds,arrows,arrows.new}

\usepackage{floatflt}
\usepackage{ltl}

\usepackage{caption}
\usepackage{subcaption}
\usepackage{multicol}

\usepackage{centernot}
\usepackage{changepage}
\usepackage{url}

\title{Causality-based Model Checking \thanks{This research was
    supported in part by the European Research Council (ERC) Grant OSARES (No. 683300), by the German Research Foundation (DFG) as part of SFB/TR 14 AVACS, and by
    the Austrian Science Fund (FWF) under
    grants S11402-N23 (RiSE and SHiNE) and Z211-N23 (Wittgenstein Award).}
}
\author{Bernd Finkbeiner
\institute{Universit\"at des Saarlandes\\
Saarbr\"ucken, Germany}
\email{finkbeiner@cs.uni-saarland.de}
\and
Andrey Kupriyanov
\institute{Institute of Science and Technology Austria\\
Klosterneuburg, Austria}
\email{andrey.kupriyanov@ist.ac.at}
}
\def\titlerunning{Causality-based Model Checking}
\def\authorrunning{B. Finkbeiner \& A. Kupriyanov}
\begin{document}
\maketitle

%\input{macros-formatting}
% General mathematical notation

\newcommand\PP{{\cal P}}

\newcommand{\mathset}[2]{\{\: #1 \;|\; #2 \:\}}
\newcommand{\bigmathset}[2]{\big\{\: #1 \;|\; #2 \:\big\}}
\newcommand{\mset}[1]{\{ #1 \}}
\newcommand{\bigmset}[1]{\big\{ #1 \big\}}

\newcommand{\ra}{\rightarrow}
\newcommand{\la}{\leftarrow}

\newcommand{\xra}[1]{\xrightarrow{#1}}
\newcommand{\xla}[1]{\xleftarrow{#1}}

\newcommand{\impl}{\ensuremath{\!\!\implies\!\!}}

\newcommand{\defeq}{\ensuremath{\triangleq}}

\newcommand{\Lang}{\ensuremath{\mathcal{L}}}

\newcommand{\tuple}[1]{\langle #1 \rangle}
\newcommand{\bigtuple}[1]{\big\langle #1 \big\rangle}

% Assertion Language

\newcommand{\StPred}{\ensuremath{\Phi(\V)}}
\newcommand{\TrPred}{\ensuremath{\Phi(\V \cup \V')}}

\newcommand{\pres}[1]{\mathit{pres}(#1)}
\newcommand{\sat}[1]{\mathit{sat}(#1)}
\newcommand{\unsat}[1]{\mathit{unsat}(#1)}
\newcommand{\ucore}[1]{\mathit{unsat\_core}(#1)}
\newcommand{\itp}[2]{\mathit{interpolate}(#1\:;\:#2)}
\newcommand{\itpB}[2]{\mathit{interpolate}\big(#1\:;\:#2\big)}

\newcommand{\var}[1]{\ensuremath{\mathit{#1}}}
\newcommand{\bvar}[1]{\ensuremath{\mathbf{#1}}}

\newcommand{\rank}[1]{\mathit{rank}(#1)}

% compleixty
\newcommand{\GI}{\ensuremath{\mathbf{GI}}}
\newcommand{\NP}{\ensuremath{\mathbf{NP}}}
\newcommand{\PSPACE}{\ensuremath{\mathbf{PSPACE}}}
\newcommand{\POL}{\ensuremath{\mathbf{P}}}
\newcommand{\EXP}{\ensuremath{\mathbf{EXP}}}

\newtheorem{assertion}{Assertion}

% Transition system
\newcommand{\Sys}{\ensuremath{S}}
% All Variables (both flexible and rigid)
\newcommand{\V}{\ensuremath{\mathcal{V}}}
% Flexible Variables
\newcommand{\Var}{\ensuremath{V}}
% Rigid Variables
\newcommand{\Vr}{\ensuremath{U}}
% Transitions
\newcommand{\T}{\ensuremath{T}}
% Initial transition
\newcommand{\Init}{\ensuremath{\Theta}}
% Transition system tuple
\newcommand{\SysT}{\ensuremath{\Sys = \langle \Var, \T, \Init \rangle}}
% Transition system language
\newcommand{\SysL}{\ensuremath{\Lang(\Sys)}}

% Transition system tuple promed
\newcommand{\SysTp}{\ensuremath{\Sys' = \langle \Var', \T', \Init' \rangle}}

% transition
\newcommand{\tr}{\ensuremath{t}}
% state
\newcommand{\st}{\ensuremath{s}}

% run
\newcommand{\run}{\ensuremath{\pi}}
\newcommand{\Run}{\ensuremath{\st_0, \st_1, \st_2, \ldots}}
\newcommand{\RunN}{\ensuremath{\st_0, \ldots, \st_{n}}}

% replace all "run" to "computation", then remove "run"
\newcommand{\comp}{\ensuremath{\pi}}
\newcommand{\Comp}{\ensuremath{\st_0, \st_1, \st_2, \ldots}}
\newcommand{\CompN}{\ensuremath{\st_0, \ldots, \st_{n}}}
\newcommand{\compN}{\ensuremath{\comp = \CompN}}

\newcommand{\StN}{\ensuremath{\{\st_0, \st_1, \ldots, \st_n\}}}

%\newcommand{\Run}{\ensuremath{\st_0, \tr_0, \st_1, \tr_1, \st_2, \tr_2, \ldots}}
%\newcommand{\RunN}{\ensuremath{\st_0, \tr_0, \st_1, \tr_1, \st_2, \tr_2, \ldots, \tr_n, \st_{n+1}}}

% Transition system indexed i
\newcommand{\Sysi}{\ensuremath{S_i}}
% Tuple of transition system indexed i
\newcommand{\SysiT}{$\Sys_i = \langle \Var_i, \T_i, \Init_i  \rangle$}

% Synchronized transition system
\newcommand{\SSys}{\ensuremath{\mathbb{S}}}
% Synchronized transitions
\newcommand{\ST}{\ensuremath{\mathbb{T}}}
% Synchronization mechanism
\newcommand{\SM}{\ensuremath{\Delta}}
% Synchronized transition system tuple
\newcommand{\SSysT}{\ensuremath{\SSys = \langle {\Sys}_1,\ldots,{\Sys}_n, \ST, \SM \rangle}}

\newcommand{\Sinput}{\!\Rightarrow\!}
\newcommand{\Soutput}{\!\Leftarrow\!}

\tikzstyle{edgelab}=[above,font=\small]

\tikzset{
    ncbar angle/.initial=90,
    ncbar/.style={
        to path=(\tikztostart)
        -- ($(\tikztostart)!#1!\pgfkeysvalueof{/tikz/ncbar angle}:(\tikztotarget)$)
        -- ($(\tikztotarget)!($(\tikztostart)!#1!\pgfkeysvalueof{/tikz/ncbar angle}:(\tikztotarget)$)!\pgfkeysvalueof{/tikz/ncbar angle}:(\tikztostart)$)
        -- (\tikztotarget)
    },
    ncbar/.default=0.5cm,
}

\tikzset{square left brace/.style={ncbar=1mm}}
\tikzset{square right brace/.style={ncbar=-1mm}}

\tikzset{round left paren/.style={ncbar=0.5cm,out=120,in=-120}}
\tikzset{round right paren/.style={ncbar=0.5cm,out=60,in=-60}}

% Concurrent trace
\newcommand{\Tr}{\ensuremath{F}}
% Events
\newcommand{\Ev}{\ensuremath{\mathcal{E}}}
% Causal links
\newcommand{\Ca}{\ensuremath{\mathcal{C}}}
% causality relation
\newcommand{\ca}{\ensuremath{\preceq}}
% Conflict relation
\newcommand{\co}{\ensuremath{\,\lightning\,}}
\newcommand{\cok}[1]{\ensuremath{\,\lightning_{\!#1}\,}}
% Event labelling
\newcommand{\Le}{\ensuremath{\lambda_\Ev}}
\newcommand{\Lek}[1]{\ensuremath{\lambda_{\Ev_#1}}}
% Causal link labelling
\newcommand{\Lc}{\ensuremath{\lambda_\Ca}}
\newcommand{\Lck}[1]{\ensuremath{\lambda_{\Ca_#1}}}
% Concurrent trace tuple
\newcommand{\TrT}{\ensuremath{\Tr = \langle \Ev, \Ca , \co ,\Le, \Lc \rangle}}
% Concurrent trace tuple only
\newcommand{\TrTo}{\ensuremath{\langle \Ev, \Ca , \co ,\Le,\Lc \rangle}}
% Concurrent trace set
\newcommand{\TrS}{\ensuremath{\mathcal{\Tr}}}
% Concurrent trace language
\newcommand{\TrL}{\ensuremath{\Lang(\Tr)}}
% entry event
\newcommand{\en}{\ensuremath{e_\triangleleft}}
% exit event
\newcommand{\ex}{\ensuremath{e_\triangleright}}

% source of a causal link
\newcommand{\src}{\ensuremath{\mathit{src}}}
% target of a causal link
\newcommand{\tgt}{\ensuremath{\mathit{tgt}}}

% Linear and compact traces
\newcommand{\Trl}{\ensuremath{\Tr}}
% Concurrent trace primed
\newcommand{\Trlp}{\ensuremath{\Trp}}
% Linear trace set
\newcommand{\TrlS}{\ensuremath{\TrS_L}}
% Compact trace set
\newcommand{\TrcS}{\ensuremath{\TrS_C}}

% Conflict relation primed
\newcommand{\cop}{\ensuremath{\:\lightning'\:}}
% Conflict relation pre-primed
\newcommand{\copp}{\ensuremath{\:{}'\lightning\:}}

% Concurrent trace primed
\newcommand{\Trp}{\ensuremath{\Tr'}}
% Concurrent trace tuple primed
\newcommand{\TrTp}{\ensuremath{\Tr' = \langle \Ev', \Ca' , \cop ,\Le' ,\Lc' \rangle}}
% Concurrent trace tuple only primed
\newcommand{\TrTop}{\ensuremath{\langle \Ev', \Ca' , \cop ,\Le' ,\Lc' \rangle}}

% Concurrent trace tuple only pre-primed
\newcommand{\TrTopp}{\ensuremath{\langle {}'\Ev, {}'\Ca, \copp ,{}'\Le ,{}'\Lc \rangle}}

% Events, indexed i
\newcommand{\Evi}{\ensuremath{\mathcal{E}_i}}
% Causal links, indexed i
\newcommand{\Cai}{\ensuremath{\mathcal{C}_i}}
% causality relation, indexed i
\newcommand{\cai}{\ensuremath{\preceq_i}}
% Conflict relation, indexed i
\newcommand{\coi}{\ensuremath{\,\lightning_i\,}}
% Conflict relation primed, indexed i
\newcommand{\copi}{\ensuremath{\:\lightning_i'\:}}
% Event labelling, indexed i
\newcommand{\Lei}{\ensuremath{\lambda_{\Ev i}}}
% Causal link labelling, indexed i
\newcommand{\Lci}{\ensuremath{\lambda_{\Ca i}}}

% Concurrent trace tuple only primed, indexed by i
\newcommand{\TrTopi}{\ensuremath{\langle \Evi', \Cai' , \copi ,\Lei' ,\Lci' \rangle}}

% Concurrent trace language primed
\newcommand{\TrLp}{\ensuremath{\Lang(\Trp)}}

% Trace morphism
\newcommand{\Trm}{\ensuremath{\mu}}
\newcommand{\Trme}{\ensuremath{\mu_\Ev}}
\newcommand{\Trmc}{\ensuremath{\mu_\Ca}}
\newcommand{\TrmT}{\ensuremath{\Trm = \langle \Trme: \Ev \ra \Ev', \Trmc: \Ca \ra \Ca' \rangle}}

% Trace run
\newcommand{\Trr}{\ensuremath{\sigma: \Ev \ra \{\CompN\}}}
\newcommand{\Trrk}[1]{\ensuremath{\sigma_#1: \Ev_#1 \ra \{\CompN\}}}

% Infinite Trace
\newcommand{\TrI}{\ensuremath{I}}
% Stem
\newcommand{\stm}{}
\newcommand{\TrIs}{\ensuremath{\Tr\stm}}
% Cycle
\newcommand{\cyc}{_\omega}
\newcommand{\TrIc}{\ensuremath{\Tr\cyc}}
% Infinite trace tuple
\newcommand{\TrIT}{\ensuremath{\TrI = \tuple{\TrIs, \TrIc}}}
% Infinite trace set
\newcommand{\TrIS}{\ensuremath{\mathcal{I}}}
% Concurrent trace language
\newcommand{\TrIL}{\ensuremath{\Lang(\TrI)}}
\newcommand{\TrIsL}{\ensuremath{\Lang(\TrIs)}}
\newcommand{\TrIcL}{\ensuremath{\Lang(\TrIc)}}

% Infinite trace stem tuple
\newcommand{\TrIsT}{\ensuremath{\TrIs = \langle \Ev, \Ca , \co ,\Le,\Lc \rangle}}

% Cycle events
\newcommand{\Evc}{\ensuremath{\mathcal{E}_\omega}}
% Cycle causal links
\newcommand{\Cac}{\ensuremath{\mathcal{C}_\omega}}
% Cycle causality relation
\newcommand{\cac}{\ensuremath{\preceq}_\omega}
% Cycle conflict relation
\newcommand{\coc}{\ensuremath{\,\lightning_\omega\,}}
% Cycle conflict relation primed
\newcommand{\cocp}{\ensuremath{\,\lightning_\omega'\,}}
\newcommand{\cocpp}{\ensuremath{\,\lightning_\omega''\,}}
% cycle event labelling
\newcommand{\Lec}{\ensuremath{\lambda_{\Ev\omega}}}
% cycle causal link labelling
\newcommand{\Lcc}{\ensuremath{\lambda_{\Ca\omega}}}

% cycle entry event
\newcommand{\enc}{\ensuremath{e_{\triangleleft\omega}}}
% cycle exit event
\newcommand{\exc}{\ensuremath{e_{\triangleright\omega}}}

% Infinite trace cycle tuple
\newcommand{\TrIcT}{\ensuremath{\TrIc = \langle \Evc, \Cac , \coc ,\Lec,\Lcc \rangle}}

% Infinite trace cycle tuple only
\newcommand{\TrIcTo}{\ensuremath{\langle \Evc, \Cac , \coc ,\Lec,\Lcc \rangle}}

% Infinite trace primed
\newcommand{\TrIp}{\ensuremath{\TrI'}}
% Infinite trace tuple primed
\newcommand{\TrITp}{\ensuremath{\TrI' = \tuple{\TrIs' , \TrIc'}}}
% Infinite trace tuple only primed
\newcommand{\TrITop}{\ensuremath{\tuple{\TrIs', \TrIc'}}}
% Infinite trace language primed
\newcommand{\TrILp}{\ensuremath{\Lang(\TrIp)}}
% stem primed
\newcommand{\TrIsp}{\ensuremath{\TrIs'}}
% cycle primed
\newcommand{\TrIcp}{\ensuremath{\TrIc'}}

% Infinite trace morphism
\newcommand{\TrIm}{\ensuremath{\nu}}
\newcommand{\TrIms}{\ensuremath{\mu\stm}}
\newcommand{\TrImc}{\ensuremath{\mu\cyc}}
\newcommand{\TrImT}{\ensuremath{\TrIm = \tuple{\TrIms,\TrImc}}}

\newcommand{\stem}{\ensuremath{\var{stem}}}
\newcommand{\cycle}{\ensuremath{\var{cycle}}}

%TIKZ

\tikzset{x=1mm,y=1mm}

\tikzstyle{lab}=[circle,draw,fill=white,inner sep=1mm, minimum height=3mm, text height=2mm]
\tikzstyle{lab'}=[circle,draw,fill=white,inner sep=0.6mm, minimum height=3mm, text height=2mm]

\tikzstyle{box}=[align=center,rectangle,draw,inner sep=0.6mm, minimum height=6mm, minimum width=6mm,text height=1.5ex,text depth=.25ex]

\tikzstyle{trace}=[framed,background rectangle/.style={draw,rounded corners},inner frame xsep=3mm, inner frame ysep=4mm, >=triangle 45, anchor=center]

\tikzstyle{tighttrace}=[framed,background rectangle/.style={draw,rounded corners},inner frame xsep=2mm, inner frame ysep=2mm, >=triangle 45, anchor=center]

\tikzstyle{conflict}=[decorate,decoration={snake,segment length=1mm,amplitude=0.5mm}]

\newcommand{\conflict}[3][]{\path (#2) edge [-,conflict,#1] node[inner sep=0mm] {$\pmb{\parallel}$} (#3)}

\newcommand{\conflictrotate}[3][]{\path (#2) edge [-,conflict,#1] node[inner sep=0mm,rotate=90] {$\pmb{\parallel}$} (#3)}

\newcommand{\lab}[2][]{
  \node[lab,#1] (#2)  {$#2$}
}

\renewcommand{\event}[3][]{
  \node[box,#1] (#2)  {$#3$}
}

\newcommand{\labevent}[3][]{
  \node[box,#1] (#2)  {$#3$};
  \node(#2lab) at(#2.north west) [lab', xshift=-0.5mm,yshift=0.5mm]  {$#2$}
}

\newcommand{\lablabevent}[4][]{
  \node[box,#1] (#2)  {$#4$};
  \node(#2lab) at(#2.north west) [lab', xshift=-0.5mm,yshift=0.5mm]  {$#3$}
}

\newcommand{\slablabevent}[4][]{
  \node[box,#1] (#2)  {$#4$};
  \node(#2lab) at(#2.north west) [lab', xshift=-0.5mm,yshift=0.5mm,inner sep=0mm]  {$#3$}
}

\newcommand{\slabevent}[3][]{
  \node[box,#1] (#2)  {$#3$};
  \node(#2lab) at(#2.north west) [lab', xshift=-0.5mm,yshift=1.1mm,inner sep=0mm]  {$#2$}
}

\newcommand{\lablab}[3][]{
  \node[lab,#1] (#2)  {$#3$}
}

\newcommand{\labeventphantom}[3][]{
  \node[box,#1,white] (#2)  {$#3$};
  \node at(#2.north west) [lab', xshift=-0.5mm,yshift=0.5mm,white]  {$#2$}
}

\newcommand{\trace}[1]{
\begin{tikzpicture}[trace]
#1
\end{tikzpicture}
}

\newcommand{\tighttrace}[1]{
\begin{tikzpicture}[tighttrace]
#1
\end{tikzpicture}
}

\def\centerarc[#1](#2)(#3:#4:#5)
{ \draw[#1] ($(#2)+({#5*cos(#3)},{#5*sin(#3)})$) arc (#3:#4:#5); }

\newcommand{\cycleEntry}[2][]{
  \node[inner sep=0.4mm](en) at($#2+(-2.9,0)$) {};
  \centerarc[#1]($#2+(6,0)$)(150:210:9mm)  
}

\newcommand{\cycleExit}[2][]{
  \node[inner sep=0.4mm](ex) at($#2+(3.1,0)$) {};
  \centerarc[#1]($#2+(-6,0)$)(30:-30:9mm)
  \node at($#2$)[xshift=4.5mm,yshift=4mm] {\hspace*{-2mm}$\omega$\hspace*{-2mm}}
}

% TO BE KILLED

\newcommand{\Abs}{$\Tr = \langle \Ev, \Ca , \co ,\Le,\Lc \rangle$}
\newcommand{\IAbs}{$I = \langle A_s, A_c, \phi_s, \phi_c \rangle$}

%\newcommand{\Trace}{$A = \langle N,E,M ,\nu,\eta \rangle$}

% Trace Productions
\newcommand{\Prod}{\ensuremath{p: (L \xra{r} R)}}
\newcommand{\ProdS}{\ensuremath{\Pi}}

\newcommand{\ConclZero}{\Lang\big(\tau_0^\Trm(\Tr)\big)}
\newcommand{\ConclOne}{\Lang\big(\tau_1^\Trm(\Tr)\big)}
\newcommand{\ConclTwo}{\Lang\big(\tau_2^\Trm(\Tr)\big)}
\newcommand{\Concli}{\Lang\big(\tau_i^\Trm(\Tr)\big)}

% Deterministic and Restricting Proof Rules

\newcommand{\DPR}{\var{DPR}}
\newcommand{\RPR}{\var{RPR}}
\newcommand{\EventOrdering}{\var{EventOrdering}}
\newcommand{\EventContraction}{\var{EventContraction}}

\newcommand{\Contractions}{\var{Contractions}}
\newcommand{\Linearizations}{\var{Linearizations}}
\newcommand{\Compactizations}{\var{Compactizations}}

% Trace Transformers

\newcommand{\pre}{\ensuremath{\mathit{pre}}}
\newcommand{\post}{\ensuremath{\mathit{post}}}

\newcommand{\Contradiction}{\ensuremath{\mathit{Contradiction}}}
\newcommand{\ContradictionP}[1]{\ensuremath{\mathit{Contradiction}(#1)}}

\newcommand{\OrderSplit}{\ensuremath{\mathit{OrderSplit}}}
\newcommand{\OrderSplitP}[1]{\ensuremath{\mathit{OrderSplit}(#1)}}
\newcommand{\OrderSplitI}{\ensuremath{\mathit{OrderSplit}^\omega}}
\newcommand{\OrderSplitIP}[1]{\ensuremath{\mathit{OrderSplit}^\omega(#1)}}

\newcommand{\EventSplit}{\ensuremath{\mathit{EventSplit}}}
\newcommand{\EventSplitP}[1]{\ensuremath{\mathit{EventSplit}(#1)}}
\newcommand{\EventSplitI}{\ensuremath{\mathit{EventSplit}^\omega}}
\newcommand{\EventSplitIP}[1]{\ensuremath{\mathit{EventSplit}^\omega(#1)}}

\newcommand{\ConflictSplit}{\ensuremath{\mathit{ConflictSplit}}}
\newcommand{\ConflictSplitP}[1]{\ensuremath{\mathit{ConflictSplit}(#1)}}

\newcommand{\Conflict}{\ensuremath{\mathit{Conflict}}}
\newcommand{\ConflictP}[1]{\ensuremath{\mathit{Conflict}(#1)}}

\newcommand{\EventRestriction}{\ensuremath{\mathit{EventRestriction}}}
\newcommand{\EventRestrictionP}[1]{\ensuremath{\mathit{EventRestriction}(#1)}}

\newcommand{\LinkRestriction}{\ensuremath{\mathit{LinkRestriction}}}
\newcommand{\LinkRestrictionP}[1]{\ensuremath{\mathit{LinkRestriction}(#1)}}

\newcommand{\CausalTrans}{\ensuremath{\mathit{CausalTransitivity}}}
\newcommand{\CausalTransP}[1]{\ensuremath{\mathit{CausalTransitivity}(#1)}}

\newcommand{\ConflictTrans}{\ensuremath{\mathit{ConflictTransitivity}}}
\newcommand{\ConflictTransP}[1]{\ensuremath{\mathit{ConflictTransitivity}(#1)}}

\newcommand{\NecessaryEvent}{\ensuremath{\mathit{NecessaryEvent}}}
\newcommand{\NecessaryEventP}[1]{\ensuremath{\mathit{NecessaryEvent}(#1)}}

\newcommand{\FirstNecessaryEvent}{\ensuremath{\mathit{FirstNecessaryEvent}}}
\newcommand{\FirstNecessaryEventP}[1]{\ensuremath{\mathit{FirstNecessaryEvent}(#1)}}

\newcommand{\LastNecessaryEvent}{\ensuremath{\mathit{LastNecessaryEvent}}}
\newcommand{\LastNecessaryEventP}[1]{\ensuremath{\mathit{LastNecessaryEvent}(#1)}}

\newcommand{\FirstLastNecessaryEvent}{\ensuremath{\mathit{(First/Last)NecessaryEvent}}}
\newcommand{\FirstLastNecessaryEventP}[1]{\ensuremath{\mathit{(First/Last)NecessaryEvent}(#1)}}

\newcommand{\ForwardUnrolling}{\ensuremath{\mathit{ForwardUnrolling}}}
\newcommand{\ForwardUnrollingP}[1]{\ensuremath{\mathit{ForwardUnrolling}(#1)}}

\newcommand{\BackwardUnrolling}{\ensuremath{\mathit{BackwardUnrolling}}}
\newcommand{\BackwardUnrollingP}[1]{\ensuremath{\mathit{BackwardUnrolling}(#1)}}

\newcommand{\Instantiate}{\ensuremath{\mathit{Instantiate}}}
\newcommand{\InstantiateP}[1]{\ensuremath{\mathit{Instantiate}(#1)}}
\newcommand{\InstantiateI}{\ensuremath{\mathit{Instantiate}^\omega}}
\newcommand{\InstantiateIP}[1]{\ensuremath{\mathit{Instantiate}^\omega(#1)}}

\newcommand{\InvarianceSplit}{\ensuremath{\mathit{InvarianceSplit}}}
\newcommand{\InvarianceSplitP}[1]{\ensuremath{\mathit{InvarianceSplit}(#1)}}

\newcommand{\NecessaryCycleEvent}{\ensuremath{\mathit{NecessaryCycleEvent}}}
\newcommand{\NecessaryCycleEventP}[1]{\ensuremath{\mathit{NecessaryCycleEvent}(#1)}}

\newcommand{\CycleToStem}{\ensuremath{\mathit{CycleToStem}}}
\newcommand{\CycleToStemP}[1]{\ensuremath{\mathit{CycleToStem}(#1)}}

\newcommand{\CycleUnrolling}{\ensuremath{\mathit{CycleUnrolling}}}
\newcommand{\CycleUnrollingP}[1]{\ensuremath{\mathit{CycleUnrolling}(#1)}}

\newcommand{\WeakFairness}{\ensuremath{\mathit{WeakFairness}}}
\newcommand{\WeakFairnessP}[1]{\ensuremath{\mathit{WeakFairness}(#1)}}

\newcommand{\StrongFairness}{\ensuremath{\mathit{StrongFairness}}}
\newcommand{\StrongFairnessP}[1]{\ensuremath{\mathit{StrongFairness}(#1)}}

% TIKZ

\newcommand{\ruleOne}[2]{
\begin{tikzpicture}
  \node(L){\trace{#1}};
  \node at(L.north)[yshift=1mm] {$L$};
  \node(R)[right=10 of L]{\trace{#2}};
  \node at(R.north)[yshift=1mm] {$R$};
  \node at($0.5*(L.east)+0.5*(R.west)$) {$\boldsymbol{\Longrightarrow}$};  
\end{tikzpicture}
}

\newcommand{\ruleOneTight}[2]{
\begin{tikzpicture}
  \node(L){\tighttrace{#1}};
  \node at(L.north)[yshift=1mm] {$L$};
  \node(R)[right=8 of L]{\tighttrace{#2}};
  \node at(R.north)[yshift=1mm] {$R$};
  \node at($0.5*(L.east)+0.5*(R.west)$) {$\boldsymbol{\Longrightarrow}$};  
\end{tikzpicture}
}

\newcommand{\ruleTwo}[3]{
\begin{tikzpicture}
  \node(L){\trace{#1}};
  \node at(L.north)[yshift=1mm] {$L$};
  \node(R1)[right=10 of L]{\trace{#2}};
  \node at(R1.north)[yshift=1mm] {$R_1$};
  \node at($0.5*(L.east)+0.5*(R1.west)$) {$\boldsymbol{\Longrightarrow}$};  
  \node(R2)[right=2 of R1]{\trace{#3}};
  \node at(R2.north)[yshift=1mm] {$R_2$};
%  \node at($0.5*(R1)+0.5*(R2)$) {$\boldsymbol{\mid}$};  
\end{tikzpicture}
}

\newcommand{\ruleTwoTight}[3]{
\begin{tikzpicture}
  \node(L){\tighttrace{#1}};
  \node at(L.north)[yshift=1mm] {$L$};
  \node(R1)[right=7 of L]{\tighttrace{#2}};
  \node at(R1.north)[yshift=1mm] {$R_1$};
  \node at($0.5*(L.east)+0.5*(R1.west)$) {$\boldsymbol{\Longrightarrow}$};  
  \node(R2)[right=1 of R1]{\tighttrace{#3}};
  \node at(R2.north)[yshift=1mm] {$R_2$};
%  \node at($0.5*(R1)+0.5*(R2)$) {$\boldsymbol{\mid}$};  
\end{tikzpicture}
}

\newcommand{\ruleMany}[4]{
\begin{tikzpicture}
  \node(L){\trace{#1}};
  \node at(L.north)[yshift=1mm] {$L$};
  \node(R0)[right=10 of L]{\trace{#2}};
  \node at(R0.north)[yshift=1mm] {$R_0$};
  \node at($0.5*(L.east)+0.5*(R0.west)$) {$\boldsymbol{\Longrightarrow}$};  
  \node(R1)[right=2 of R0]{\trace{#3}};
  \node at(R1.north)[yshift=1mm] {$R_1$};
  \node(Rk)[right=5 of R1]{\trace{#4}};
  \node at(Rk.north)[yshift=1mm] {$R_k$};
  \node at($0.5*(R1.east)+0.5*(Rk.west)$) {$\boldsymbol{\ldots}$};  
\end{tikzpicture}
}

\newcommand{\ruleManyTwo}[3]{
\begin{tikzpicture}
  \node(L){\tighttrace{#1}};
  \node at(L.north)[yshift=1mm] {$L$};
  \node(R1)[right=8 of L]{\tighttrace{#2}};
  \node at(R1.north)[yshift=1mm] {$R_1$};
  \node at($0.5*(L.east)+0.5*(R1.west)$) {$\boldsymbol{\Longrightarrow}$};  
  \node(Rk)[right=5 of R1]{\tighttrace{#3}};
  \node at(Rk.north)[yshift=1mm] {$R_k$};
  \node at($0.5*(R1.east)+0.5*(Rk.west)$) {$\boldsymbol{\ldots}$};  
\end{tikzpicture}
}

\newcommand{\ruleTwoConcl}[4]{
\begin{tikzpicture}
  \node(R1){\trace{#1}};
  \node at(L.north)[yshift=1mm] {$#2$};
  \node(R2)[right=10 of L]{\trace{#3}};
  \node at(R.north)[yshift=1mm] {$#4$};
\end{tikzpicture}
}

% Trace Unwinding

\newcommand{\Unw}{\ensuremath{\Upsilon}}

% nodes, internal nodes, leaves
\newcommand{\N}{\ensuremath{N}}
\newcommand{\Ni}{\ensuremath{N_I}}
\newcommand{\Nl}{\ensuremath{N_L}}
\newcommand{\Nr}{\ensuremath{N_R}}
% edges
\newcommand{\F}{\ensuremath{E}}
% labelling: forest edges -> trace productions
\newcommand{\Lf}{\ensuremath{\delta}}
% labelling: nodes -> traces
\newcommand{\Ln}{\ensuremath{\gamma}}
\newcommand{\LnLang}[1]{\Lang\big(\Ln(#1)\big)}

% labelling: internal nodes -> trace morphisms (premise -> concrete trace)
\newcommand{\Li}{\ensuremath{\mu}}

\newcommand{\UnwT}{\ensuremath{\Unw = \tuple{\N, \F, \Ln, \Lf, \Li }}}

\newcommand{\Satisfiable}{\ensuremath{\mathit{Satisfiable}}}
\newcommand{\SatisfiableP}[1]{\ensuremath{\mathit{Satisfiable}(#1)}}
\newcommand{\SatisfiableBP}[1]{\ensuremath{\mathit{Satisfiable}\big(#1\big)}}

\newcommand{\UnsatSubtrace}{\ensuremath{\mathit{UnsatSubtrace}}}
\newcommand{\UnsatSubtraceP}[1]{\ensuremath{\mathit{UnsatSubtrace}(#1)}}
\newcommand{\UnsatSubtraceBP}[1]{\ensuremath{\mathit{UnsatSubtrace}\big(#1\big)}}

\newcommand{\SSA}{\ensuremath{\mathit{SSA}}}

\newcommand{\Abstract}{\ensuremath{\mathit{Abstract}}}
\newcommand{\AbstractSP}{\ensuremath{\Abstract(\Sys,\varphi)}}

\newcommand{\assign}{\ensuremath{\longleftarrow}}

\newcommand{\Refine}{\ensuremath{\mathit{SafetyRefinement}}}
\newcommand{\RefineP}[1]{\ensuremath{\mathit{SafetyRefinement}(#1)}}

\newcommand{\LivenessRefine}{\ensuremath{\mathit{LivenessRefinement}}}
\newcommand{\LivenessRefineP}[1]{\ensuremath{\mathit{LivenessRefinement}(#1)}}

\newcommand{\ApplyRule}{\ensuremath{\mathit{Apply}}}
\newcommand{\ApplyRuleP}[1]{\ensuremath{\mathit{Apply}(#1)}}

\newcommand{\InitialAbstraction}{\ensuremath{\mathit{InitialAbstraction}}}
\newcommand{\InitialAbstractionP}[1]{\ensuremath{\mathit{InitialAbstraction}(#1)}}

\newcommand{\InitialAbstractTableau}{\ensuremath{\mathit{InitialAbstractTableau}}}
\newcommand{\InitialAbstractTableauP}[1]{\ensuremath{\mathit{InitialAbstractTableau}(#1)}}

\newcommand{\Terminating}{\ensuremath{\mathit{Terminating}}}
\newcommand{\TerminatingP}[1]{\ensuremath{\mathit{Terminating}(#1)}}
\newcommand{\TerminatingBP}[1]{\ensuremath{\mathit{Terminating}\big(#1\big)}}

\tikzstyle{arr}=[->,double,double distance=0.5mm,>=stealth new, arrow head=2mm]

% Trace Tableau

\newcommand{\Tab}{\ensuremath{\Gamma}}

\newcommand{\Cov}{\ensuremath{\leadsto}}

\newcommand{\TabT}{\ensuremath{\Tab = \tuple{\N, \F, \Ln, \Lf, \Li, \Cov}}}

% Abstract Trace Tableau

\newcommand{\Tabac}{\ensuremath{\Delta}}

% labelling: nodes -> abstract traces
\newcommand{\Lna}{\ensuremath{\hat{\gamma}}}
% labelling: edges -> abstract productions
\newcommand{\Lfa}{\ensuremath{\hat{\delta}}}
% labelling: internal nodes -> trace morphisms (premise -> abstract trace)
\newcommand{\Lia}{\ensuremath{\hat{\mu}}}
% labelling: nodes -> trace morphisms (abstract trace -> concrete trace)
\newcommand{\Lac}{\ensuremath{\sigma}}

\newcommand{\TabacT}{\ensuremath{\Tabac = \tuple{\N, \F, \Ln, \Lf, \Li, \Cov, \Lna, \Lfa, \Lia, \Lac}}}

\newcommand{\Taba}{\ensuremath{\hat{\Gamma}}}

\newcommand{\TabaT}{\ensuremath{\Taba = \tuple{\N, \F, \Lna, \Lfa, \Lia, \Cov}}}

\setlength{\algomargin}{0.2em}
\SetAlFnt{\small}
\DontPrintSemicolon

\newcommand{\tpre}{\pre(\tau)}
\newcommand{\mh}{{\hat{m}}}

% LTL

% safety property
\newcommand{\SafeProp}{\ensuremath{\varphi}}

\newcommand{\sep}{\:\:\big|\:\:}

\newcommand{\Next}{\ensuremath{\LTLcircle}}
\newcommand{\Globally}{\ensuremath{\LTLsquare}}
\newcommand{\Finally}{\ensuremath{\LTLdiamond}}
\newcommand{\Until}{\ensuremath{\LTLuntil}}
\newcommand{\Release}{\ensuremath{\LTLrelease}}
\newcommand{\Previous}{\ensuremath{\LTLprevious}}
\newcommand{\Once}{\ensuremath{\LTLpastfinally}}
\newcommand{\StrictOnce}{\ensuremath{\widehat{\:\Once\:}}}
\newcommand{\Weakuntil}{\ensuremath{\LTLweakuntil}}

\begin{abstract}
Model checking is usually based on a comprehensive traversal of the state space. Causality-based model checking is a radically different approach that instead analyzes the cause-effect relationships in a program. We give an overview on a new class of model checking algorithms that capture the causal relationships in a special data structure called concurrent traces. Concurrent traces identify key events in an execution history and link them through their cause-effect relationships. The model checker builds a tableau of concurrent traces, where the case splits represent different causal explanations of a hypothetical error. Causality-based model checking has been implemented in the \textsc{Arctor} tool, and applied to previously intractable multi-threaded benchmarks.
\end{abstract}

\section{Introduction}

%Concurrency fascinated many generations of computer scientists -- the intriguing interdependence between concurrent events, and how they cooperate to achieve a common goal. A concurrent program can do many things \emph{in parallel}, with its components either physically or logically distributed, thus achieving the processing speed unattainable by sequential programs. And precisely this power gives rise to problems: many more things can also go wrong.

Model checking \cite{Clarke/Design-and-Synthesis} is a ``push-button'' technology: it verifies a given program completely automatically, without any help from the human user. A fundamental challenge, however, is the infamous state space explosion problem: in a concurrent program, the number of control states grows exponentially with the number of parallel threads. This creates a barrier for the standard approach to model checking, which is based on a comprehensive traversal of the state space.

%Verifying program properties is an important task; there exist plenty of manual and automatic approaches that either try to catch bugs or to prove that a given program satisfies its specification, and, therefore, will serve its intended design purpose. Among verification methods, \emph{model checking} attempts to construct a proof or to find a counterexample completely automatically -- it provides the system designer with an appealing ``push-button'' approach to program verification. While very successful for sequential programs, model checking of concurrent programs suffers from the so-called state space explosion problem: the number of control states of a concurrent program grows exponentially fast with the addition of new components. The problem is exaggerated with large data domains for program variables.

Causality-based model checking~\cite{Safety,Liveness,Thesis} is inspired by the observation that it is not unusual for programs that are difficult to verify with a conventional model checker to have surprisingly short ``paper-and-pencil'' proofs.
%in many cases a human is able to devise either a short proof of correctness, or a small counterexample; this holds even for large systems, which are out of reach for any automatic method. We claim that this happens
A likely explanation is that humans reason in terms of \emph{causal relationships}. In most general terms, causality can be defined as the relation between two events, where the first event (the \emph{cause}) is understood to be partly responsible for the second (the \emph{effect}). Reasoning by causality is the usual style of constructing proofs: assume some situation (the effect) to be present, and derive all possible explanations (the causes). Consider the following assertion and its proof from Leslie Lamport's paper \cite{Lamport/Bakery} introducing the Bakery algorithm for mutual exclusion:

\begin{adjustwidth}{5mm}{5mm}
%\begin{quote}
\begin{assertion} If processors $i$ and $k$ are in the bakery and $i$ entered the bakery before $k$ entered the doorway, then $\var{number}[i]<\var{number}[k]$.
\end{assertion}

%\vspace*{-3mm}
\begin{proof}
By hypothesis, $\var{number}[i]$ had its current value while $k$ was choosing the current value of $\var{number}[k]$. Hence, $k$ must have chosen $\var{number}[k] \ge 1 + \var{number}[i]$.
\end{proof}
%\end{quote}
\end{adjustwidth}

The proof starts by assuming the situation where the event ``$i$ entered the bakery'' precedes the event ``$k$ entered the doorway'', and, moreover, $\var{number}[i]$ preserves its value between two events. The proof proceeds by deriving from this situation another \emph{necessary} fact (notice the words ``\textit{must have chosen}''): $(\var{number}[k] \ge 1 + \var{number}[i])$.

 Unlike standard model checking, causality-based model checking is not based on a traversal of the state space but instead tracks the causal dependencies in the system.
%
%The solution we describe aims to uncover the intricate interplay between dependency and independence.
In concurrent programs, it is often the case that not many concurrent events depend on each other -- most events are, in fact, independent, and precisely this allows concurrent programs to achieve better performance than sequential programs.
Causality-based model checking provides a formal proof system as well as an automatic method for constructing proofs or finding counterexamples employing the principles of causal reasoning.

%On the other hand, some dependencies do exist. Moreover, they complicate the reasoning about concurrency to such an extent that no general solution to the state space explosion problem is possible: there will always be programs requiring exponential resources for their analysis.
%
%Despite this theoretical complexity,

\section{The Causality-based Model Checking Framework}

We now introduce the main components of the causality-based model checking framework. %For a more formal presentation, we refer the reader to \cite{Safety,Liveness,Thesis}.% and motivate it by a simple example.

%\vspace*{-3mm}
\paragraph{Concurrent traces.} 
The basic building block, which we intend as a replacement of \emph{state} in the standard model checking, is called a \emph{concurrent trace}. Instead of a single momentary snapshot of the program computation, it represents a set of related computation events. Each event is labeled by a \emph{transition predicate}, and describes a set of program transitions satisfying the predicate. Events are related by \emph{causal links}, which are simply the ordering constraints between events. Causal links are also labeled by transition predicates; these constraints represent conditions which all events in scope of the causal link should satisfy. Finally, there is a \emph{conflict relation}, which prohibits certain events to coincide in time.

\begin{floatingfigure}[r]{7cm}
\scalebox{0.8}{
\tighttrace{
    \labevent{i}{x=y};
    \labevent[right=14 of i,yshift=8mm]{a}{x'=x+1};
    \labevent[right=14 of i,yshift=-8mm]{b}{y'=y-1};
    \labevent[right=50 of i]{c}{x>y};
    \path (i.east) edge[->] (a.west);
    \path (i.east) edge[->] (b.west);
    \path (a.east) edge[->] node[above,yshift=1.5mm] {$x'=x$} (c.west);
    \path (b.east) edge[->] node[below,yshift=-1mm] {$y'=y$} (c);
    \conflictrotate{a}{b};
  }
}
\end{floatingfigure}

A concurrent trace as a whole can be understood as a combination of existential and universal constraints, which collectively describe a set of program computations, such that the concurrent trace can be mapped to each of those computations respecting the constraints. E.g., a concurrent trace on the right represents the set of computations which satisfy the following formula:
%\vspace*{-2mm}
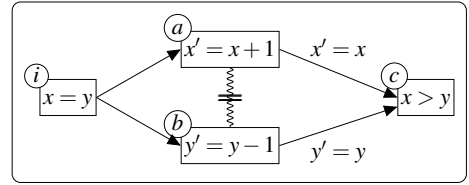
\[
\begin{array}{ll}
\exists i,a,b,c \:. \!\!\!\!\!\!& (i \le a) \:\land\:   (i \le b) \:\land\:   (a \le c) \:\land\:   (b \le c) \land (a \ne b) \:\land   \\
                        & (s_i,s_{i+1}) \models (x\!=\!y) \:\land\:  
                          (s_a,s_{a+1}) \models (x'\!=\!x\!+\!1)  \:\land\:
                          (s_b,s_{b+1}) \models (y'\!=\!y\!-\!1)  \:\land\:
                          (s_c,s_{c+1}) \models (x\!>\!y) \\
\land \: \forall  j, k . & (a < j < c) \implies (s_j,s_{j+1}) \models (x'\!=\!x) \:\land\:
                        (b < k < c) \implies (s_k,s_{k+1}) \models (y'\!=\!y) 
\end{array}
\]
%\vspace*{-2mm}

In \cite{Liveness} we have extended the basic model of \emph{finite} concurrent traces from \cite{Safety} to \emph{infinite} concurrent traces. An infinite concurrent trace consists of a \emph{stem} and a \emph{cycle}, each being a concurrent trace, with the semantics that the cycle occurs infinitely often after the stem occurs once.

\textit{Graphical Notation.} We show event identities in circles, and labeling formulas in squares.  Causal links are shown as solid lines with arrows, and conflicts as crossed zigzag lines. The cycle part of the trace is depicted in round brackets, superscripted with $\omega$. We omit any of those when they are not important or would create clutter in the current context. 

%\begin{figure}[t]
%\centering
%\begin{tikzpicture}
%  \node(l1){\textbf{loop forever do}};
%  \node(n1)[draw,rectangle,below=1 of l1]{$N_1$};
%  \node(t1)[draw,rectangle,below=1 of n1]{$T_1$};
%  \node(c1)[draw,rectangle,below=1 of t1]{$C_1$};
%
%  \node[left=15 of t1,yshift=2mm] (p1) {$P_1 ::$};
%  \node[below=22 of l1.west] (ew1){};
%  \node[below=22 of l1.east] (ee1){};
%
%  \draw ($(c1.south west)+(-3,0)$) to [square left brace ] ($(n1.north west)+(-3,0)$);
%  \draw ($(n1.north east)+(3,0)$) to [square left brace ] ($(c1.south east)+(3,0)$);
%  
%  \draw ($(ew1)+(-1,0)$) to [square left brace] ($(l1.north west)+(-1,-1)$);
%  \draw ($(l1.north east)+(1,-1)$) to [square left brace] ($(ee1)+(1,0)$);
%	
%  \node[right=25 of l1] (l1){\textbf{loop forever do}};
%  \node(n1)[draw,rectangle,below=1 of l1]{$N_2$};
%  \node(t1)[draw,rectangle,below=1 of n1]{$T_2$};
%  \node(c1)[draw,rectangle,below=1 of t1]{$C_2$};
%
%  \node[left=15 of t1,yshift=2mm] (p1) {$P_2 ::$};
%  \node[below=22 of l1.west] (ew1){};
%  \node[below=22 of l1.east] (ee1){};
%
%  \draw ($(c1.south west)+(-3,0)$) to [square left brace ] ($(n1.north west)+(-3,0)$);
%  \draw ($(n1.north east)+(3,0)$) to [square left brace ] ($(c1.south east)+(3,0)$);
%  
%  \draw ($(ew1)+(-1,0)$) to [square left brace] ($(l1.north west)+(-1,-1)$);
%  \draw ($(l1.north east)+(1,-1)$) to [square left brace] ($(ee1)+(1,0)$);
%
%  \node[left=2 of p1]{$\parallel$};
%	
%\end{tikzpicture}
%\caption{Schematic mutual exclusion program}
%\label{fig:mutual-exclusion}
%\end{figure}

%\vspace*{-3mm}
\paragraph{Representation of program properties.} Given a program property of interest, we start our analysis by representing the \emph{property violation} as a concurrent trace. From our experience, we find out that violations of many useful properties can be quite naturally expressed as concurrent traces.

%For an illustration, s
Suppose we are given a program $P$, consisting of two processes $P_1$ and $P_2$ that require mutually exclusive access to their critical sections. 
%In the schematic representation of Figure~\ref{fig:mutual-exclusion}, 
Each process $P_i$, $i=1,2$, contains an endless loop \tikz[anchor=base, baseline]{\node[outer sep=0](n){$N_i \ra T_i \ra C_i$}; \draw[->,>=latex] (n.east) .. controls ($(n.east)+(5mm,-4mm)$) and ($(n.west)+(-5mm,-4mm)$) .. (n.west);} with three sections: $N_i$ (``noncritical''), $T_i$ (``trying''), and $C_i$ (``critical''). We can easily represent violations of the desired properties of $P$ as concurrent traces (we denote the initial state of $P$ as  \Init).

\ \\
\begin{floatingfigure}[r]{3.5cm}
\scalebox{0.8}{
\tighttrace
{
  \event{a}{\Init};
  \event[right=10 of a]{b}{C_1 \land C_2};
  \path (a) edge[->] (b);
}
}
\end{floatingfigure}
\textit{Mutual exclusion.} Processes should not be in their critical sections simultaneously, which is described by the formula $\Globally  \neg(C_1 \land C_2)$.
%All violations of the property are contained in the language of the concurrent trace on the right.

\begin{floatingfigure}[r]{5.1cm}
\scalebox{0.8}{
\tighttrace
{
  \event{a}{\Init};
  \event[right=10 of a]{b}{T_1 \land \neg C_2};
  \event[right=12 of b]{c}{C_2};
  \path (a) edge[->] (b);
  \path (b) edge[->] node[above] {$\neg C_1$} (c);
}
}

\end{floatingfigure}
\textit{Strict precedence.} When $P_1$ is trying to get access and $P_2$ is not in the critical section, then $P_1$ will be admitted to the critical section first: $ \Globally \big(\: (T_1 \land \neg C_2) \implies (\neg C_2) \Weakuntil C_1 \: \big) $. %The violations are captured by the concurrent trace on the right.

\begin{floatingfigure}[r]{5.7cm}
\scalebox{0.8}{
\tighttrace
{
  \event{a}{\Init};
  \event[right=6 of a]{b}{T_1 };
  \event[right=6 of b]{c}{C_2};
  \event[right=6 of c]{d}{\neg C_2};
  \event[right=6 of d]{e}{C_2};
  \path (a) edge[->] (b);
  \path (b) edge[->] (c);
  \path (c) edge[->] (d);
  \path (d) edge[->] (e);
  \path (b) edge[->] (c);
  \path (b.north east) edge[->,bend left=22] node[above] {$\neg C_1$} (e.north west);
}
}
\vspace*{-3mm}
\end{floatingfigure}
\textit{Bounded overtaking.} We may want to give an upper bound on the amount of overtaking, where overtaking means that one process enters the critical section ahead of its rival. The property of 1-bounded overtaking can be expressed as\\ $ \Globally \big(\: T_1 \implies (\neg C_2) \Weakuntil (C_2) \Weakuntil (\neg C_2) \Weakuntil (C_1) \: \big) $.
%The violations of the property can be described by the concurrent trace on the right.

\vspace*{-3mm}
\paragraph{Trace transformers.} Our analysis proceeds in steps, where each step takes some concurrent trace, and applies to it a so called \emph{trace transformer}. Trace transformers are the proof rules describing necessary consequences from the current analysis situation represented as a concurrent trace. There is a number of general proof rules for safety and liveness analysis that were presented in our previous work \cite{Thesis,Safety,Liveness}; we also envision that application-specific proof rules can be easily created for particular domains.

A trace transformer is an ordered set of \emph{trace productions} of the form $(L \xra{r_i} R_i)$, where all productions share the same left-hand side $L$. A trace production is a generalization of a \emph{graph production} \cite{DBLP:conf/gg/1997handbook} and describes a formal rule to transform one concurrent trace into another. Collectively, all productions of a trace transformer encode some case distinction, where the union of trace languages of individual cases is contained in the language of the original trace. Below we show several examples of trace transformers.%, with their graphical representation shown to the right from the textual description.

\begin{floatingfigure}[r]{7.4cm}
\scalebox{0.8}{
\ruleTwoTight
{
  \lab{a};
  \lab[right=7 of a]{b};
}
{
  \lab{a};
  \lab[right=7 of a]{b};
  \path (a) edge[->] (b);
}
{
  \lab{a};
  \lab[right=7 of a]{b};
  \path (a) edge[<-] (b);
}
}

\end{floatingfigure}
\OrderSplitP{a,b}\ trace transformer considers alternative orderings of two concurrent events $a$ and $b$. 

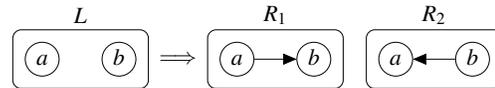
\begin{floatingfigure}[r]{9.3cm}
\vspace*{5mm}
\scalebox{0.8}{
\ruleOneTight
{
  \labevent{a}{\varphi'};
%  \labeventphantom[right=2 of a,yshift=8mm]{c}{\varphi \land \neg \varphi'};
  \labevent[right=20 of a]{b}{\neg \varphi};
  \path (a) edge[->] (b);
  \conflict[bend right]{a}{b};
}
{
  \labevent{a}{\varphi'};
  \labevent[right=10 of a,yshift=8mm]{c}{\varphi \land \neg \varphi'};
  \labevent[right=12 of c,yshift=-8mm]{b}{\neg \varphi};
  \path (a) edge[->] (b);
  \conflict[bend right=20]{a}{b};
  \path (a) edge[->,bend left=20] (c);
  \path (c.east) edge[->,bend left=16] (blab);
  \conflict{a.east}{c.south west};
  \conflict{c.south east}{b.west};
}
}
\vspace*{-6mm}
\end{floatingfigure}
\NecessaryEventP{a,b,\varphi}, given two causally related and conflicting events $a$ and $b$ in a concurrent trace, and a state predicate $\varphi \in \StPred$, such that the label of $a$ implies $\varphi'$, and the label of $b$ implies $\neg \varphi$, introduces a new ``bridging'' event $c$ in between. This condition can be interpreted as a contradiction between events $a$ and $b$ ($a$ ``ends'' in the region $\varphi$, while $b$ ``starts'' in the region $\neg\varphi$).

\begin{floatingfigure}[r]{7.8cm}
\scalebox{0.8}{
\ruleTwoTight
{
  \lab[white]{a};
  \cycleEntry{(a.west)};
  \cycleExit{(a.east)};
}
{
  \lab[white]{a};
  \lab[right=3 of a,white]{b};
  \cycleEntry{(a.west)};
  \cycleExit{(b.east)};
  \path (en) edge[->] node[above] {$\varphi$} (ex);
}
{
  \labevent{a}{\neg\varphi};
  \cycleEntry{(a.west)+(-5,0)};
  \cycleExit{(a.east)+(3,0)};
}
}
\vspace*{-3mm}
\end{floatingfigure}
\InvarianceSplitP{\varphi}\ makes a case distinction about the program behavior at infinity: for a given predicate $\varphi$ either all events in the cycle part satisfy it, or a violating event should happen infinitely often. 

Due to the space limitations we are not able to show here, even informally, all trace transformers used in the examples of the next section; the interested reader should refer to \cite{Thesis, Safety,Liveness} for their formal description.

\vspace*{-3mm}
\paragraph{Tableau construction.} The final component of our framework is a collection of datastructures for the organization of proof or counterexample search. Each datastructure comes with
the corresponding algorithm for its automatic construction; the datastructures and algortihms have different properties which are detailed in Figure~\ref{fig:tableaux}.

The conceptually simplest datastructure, called \emph{trace unwinding}, is a forest of nodes each labeled with a concurrent trace. The forest roots represent concurrent traces that encode possible violations of the property of interest. The exploration algorithm proceeds by picking some forest leaf, and employing an applicable trace transformer, producing a number of further nodes. The exploration stops when all forest leaves are found to be contradictory.

The most involved datastructure, \emph{abstract trace tableau}, contains, besides concrete trace unwinding, an abstract \emph{looping trace tableau}, which may have covering edges between tableau nodes. A covering condition is an extension of subgraph isomorphism to concurrent traces: a concurrent trace, which is a subgraph of another trace, represents a situation which was encountered in the analysis before. This tableau is also allowed to contain \emph{causal loops}, i.e. infinite repetitions of a sequence of trace productions, which together imply the impossibility of a computation satisfying them. The abstract trace tableau is used to track premises of already applied proof rules, and, thus, simplifies coverings.

\tikzstyle{ds}=[text width=6cm,rectangle,draw,align=right,font=\small]
\tikzstyle{dsp}=[text width=8cm,align=left,font=\small,yshift=2mm]

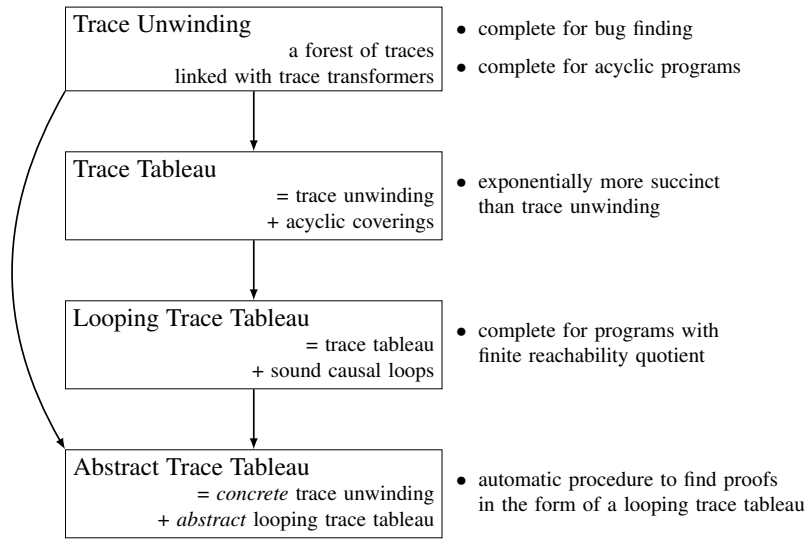
\begin{figure}[t]

\centering

\scalebox{0.8}{
\begin{tikzpicture}

\node(u)[ds]{
{\large Trace Unwinding} \hfill \phantom{a}  \\ a forest of traces \\ linked with trace transformers
};

\node[dsp,right=-5mm of u]{
\begin{itemize}
\item complete for bug finding
\item complete for acyclic programs
\end{itemize}
};

\node(t)[ds,below=1cm of u] {{\large Trace Tableau} \hfill \phantom{a}\\ = trace unwinding\\ + acyclic coverings};

\node[dsp,right=-5mm of t]{
\begin{itemize}
\item exponentially more succinct\\ than trace unwinding
\end{itemize}
};

\node(lt)[ds,below=1cm of t] {{\large Looping Trace Tableau} \hfill \phantom{a}\\ = trace tableau\\ + sound causal loops};

\node[dsp,right=-5mm of lt]{
\begin{itemize}
\item complete for programs with\\ finite reachability quotient
\end{itemize}
};

\node(at)[ds,below=1cm of lt] {{\large Abstract Trace Tableau} \hfill \phantom{a}\\ = \emph{concrete} trace unwinding\\ + \emph{abstract} looping trace tableau};

\node[dsp,right=-5mm of at]{
\begin{itemize}
\item automatic procedure to find proofs\\ in the form of a looping trace tableau 
\end{itemize}
};

\path (u) edge[->, thick,>=latex] (t);
\path (t) edge[->, thick,>=latex] (lt);
\path (lt) edge[->, thick,>=latex] (at);
\path (u.south west) edge[->, thick,>=latex, bend right] (at.north west);

\end{tikzpicture}
}
\caption{Data structures for causality-based model checking.}
\label{fig:tableaux}
\vspace*{-2mm}
\end{figure}

\section{Two Examples}

We illustrate the framework with two examples taken from \cite{Safety} and \cite{Liveness}: one for the analysis of a safety property (reachability), and another for the analysis of a liveness property (termination).

\begin{figure}[t]
\centering
\scalebox{0.8}{
\begin{tikzpicture}[>=stealth',every state/.style= {inner sep=0.7mm,minimum size=0pt}, node distance=1.2cm]

      \node[state]    (s1)  at  (1cm, 3cm) {$r_1$};
      \node (sstart) [circle,above=0.3cm of s1] {};
      \node[state,below of= s1] (s2) {$r_2$};
      \node[state, below of= s2] (s3) {$r_3$};
      \path
      (sstart)  edge [->]  (s1)
      (s1)  edge [->] node[right] {$a_0$} (s2)
      (s2)  edge [->] node[right] {$c_0$} (s3)
      ;

      \node[state]    (s1)  at  (4cm, 3cm) {$s_1$};
      \node (sstart) [circle,above=0.3cm of s1] {};
      \node[state,below of= s1] (s2) {$s_2$};
      \node[state, below of= s2] (s3) {$s_3$};
      \node[state, left of= s2] (s4) {$s_4$};
      \path
      (sstart)  edge [->]  (s1)
      (s1)  edge [->] node[right] {$b_1$} (s2)
      (s1)  edge [->] node[left] {$a_1$} (s4)
      (s2)  edge [->] node[right] {$c_1$} (s3)
      ;

      \node[state]    (s1)  at  (6cm, 3cm) {$t_1$};
      \node (sstart) [circle,above=0.3cm of s1] {};
      \node[state,below of= s1] (s2) {$t_2$};
      \node[state, below of= s2] (s3) {$t_3$};
      \path
      (sstart)  edge [->]  (s1)
      (s1)  edge [->] node[right] {$b_2$} (s2)
      (s2)  edge [->] node[right] {$c_2$} (s3)
      ;

      \node[state]    (s1)  at  (8cm, 3cm) {$u_1$};
      \node (sstart) [circle,above=0.3cm of s1] {};
      \node[state,below of= s1] (s2) {$u_2$};
      \node[state, below of= s2] (s3) {$u_3$};
      \path
      (sstart)  edge [->]  (s1)
      (s1)  edge [->] node[right] {$b_3$} (s2)
      (s2)  edge [->] node[right] {$c_3$} (s3)
      ;

      \node[right = 0.4cm of s2] {$\cdots$};

      \node[state]    (s1)  at  (10cm, 3cm) {$v_1$};
      \node (sstart) [circle,above=0.3cm of s1] {};
      \node[state,below of= s1] (s2) {$v_2$};
      \node[state, below of= s2] (s3) {$v_3$};
      \path
      (sstart)  edge [->]  (s1)
      (s1)  edge [->] node[right] {$b_n$} (s2)
      (s2)  edge [->] node[right] {$c_n$} (s3)
      ;      
       \end{tikzpicture}
}
{\small
\begin{align*}
& \ST = \bigmset{ \bvar{a} = \mset{a_0,a_1},\, 
 \bvar{b_1} = \mset{b_1},\, \ldots\, ,\, \bvar{b_n} = \mset{b_n},\,
 \bvar{c} = \mset{c_0,c_1,c_2,c_3,\ldots,c_n} }\\
& \Init \equiv r_1' \land s_1' \land t_1' \land u_1'\land \ldots \land v_1' 
\end{align*}
}

\scalebox{0.8}{
\begin{tikzpicture}[baseline={([yshift=-0.8ex]current bounding box.center)}]
  \node(L){1:\tighttrace{
    \event{i}{\Init};
    \event[right=8 of i]{c}{\bvar{c}};
    \path (i.east) edge[->] (c.west);
  }};
  \node(R2)[right=43 of L]{2:\tighttrace{
    \event{i}{\Init};
    \event[right=8 of i]{a}{\bvar{a}};
    \event[right=30 of i]{c}{\bvar{c}};
    \path (i.east) edge[->] (a.west);
    \path (a.east) edge[->] node[above] {$r_2 \land r_2'$} (c.west);
  }};
  \node(R3)[below=8 of R2]{3:\tighttrace{
    \event{i}{\Init};
    \event[right=8 of i,yshift=5mm]{a}{\bvar{a}};
    \event[right=8 of i,yshift=-5mm]{b}{\bvar{b_1}};
    \event[right=30 of i]{c}{\bvar{c}};
    \path (i.east) edge[->] (a.west);
    \path (i.east) edge[->] (b.west);
    \path (a.east) edge[->] node[above] {$r_2 \land r_2'$} (c);
    \path (b.east) edge[->] node[below] {$s_2 \land s_2'$} (c);
  }};

  \node(R4)[left=25 of R3]{4:\tighttrace{
    \event{i}{\Init};
    \event[right=8 of i]{a}{\bvar{a}};
    \event[right=10 of a]{b}{\bvar{b_1}};
    \event[right=45 of i]{c}{\bvar{c}};
    \path (i.east) edge[->] (a.west);
    \path (a) edge[->, bend left=25] node[above] {$r_2 \land r_2'$} (c);
    \path (b.east) edge[->] node[below] {$s_2 \land s_2'$} (c);
    \path (a) edge[->] (b);    
  }};
  
  \node(R6)[below=8 of R4]{6:\tighttrace{
    \event{i}{\Init};
    \event[right=8 of i]{a}{\bvar{a}};
    \event[right=20 of a]{b}{\bvar{b_1}};
    \event[right=55 of i]{c}{\bvar{c}};
    \path (i.east) edge[->] (a.west);
    \path (a) edge[->, bend left=25] node[above] {$r_2 \land r_2'$} (c);
    \path (b.east) edge[->] node[below] {$s_2 \land s_2'$} (c);
    \path (a) edge[->] (b);

    \event[right=2 of a,yshift=-7mm]{s}{\neg s_1 \land s_1'};

    \path (a) edge[->,bend right] (s);
    \path (s) edge[->,bend right] node[pos=0, right,xshift=1.5mm] {$s_1 \land s_1'$} (b);
    
  }};

   \node[below=0 of R6]{$\bot$};  

  \node(R5)[below=8 of R3,xshift=0mm]{5:\tighttrace{
    \node(i)[inner sep=0mm, outer sep=0mm]{$\cdots$};    
    \event[right=0 of i,]{b}{\bvar{b_1}};
    \event[right=8 of b]{a}{\bvar{a}};
    \node[right=1 of a,inner sep=0mm]{$\cdots$};    
    \path (b) edge[->] (a);    
  }};

  \node(R7)[below=8 of R5,xshift=0mm]{7:\tighttrace{
    \node(i)[inner sep=0mm, outer sep=0mm]{$\cdots$};    
    \event[right=0 of i,]{b}{\bvar{b_1}};
    \event[right=20 of b]{a}{\bvar{a}};
    \node[right=1 of a,inner sep=0mm]{$\cdots$};    
    \path (b) edge[->] (a);    
    
    \event[right=2 of b,yshift=-7mm]{s}{\neg s_1 \land s_1'};

    \path (b) edge[->,bend right] (s);
    \path (s) edge[->,bend right] node[pos=0, right,xshift=1.5mm] {$s_1 \land s_1'$} (a);    
  }};

   \node[below=0 of R7]{$\bot$};  

  \path (L) edge[arr] node[above] {\textit{NecessaryEvent}} (R2);
  \path (R2) edge[arr] node[right] {\textit{NecessaryEvent}} (R3);
  \path (R3) edge[arr] node[above] {\textit{OrderSplit}} (R4);
  \path (R4) edge[arr] node[right] {\textit{NecessaryEvent}} (R6);
  \path (R3) edge[arr] node[right] {\textit{OrderSplit}} (R5);
  \path (R5) edge[arr]node[right] {\textit{NecessaryEvent}}  (R7);
   
\end{tikzpicture}
}
\caption{\emph{Top}: example system from \cite{Esparza/Unfoldings}. \emph{Bottom}: its correctness proof in the form of a trace unwinding.}
\label{fig:unfoldings}
\vspace*{-2mm}
\end{figure}
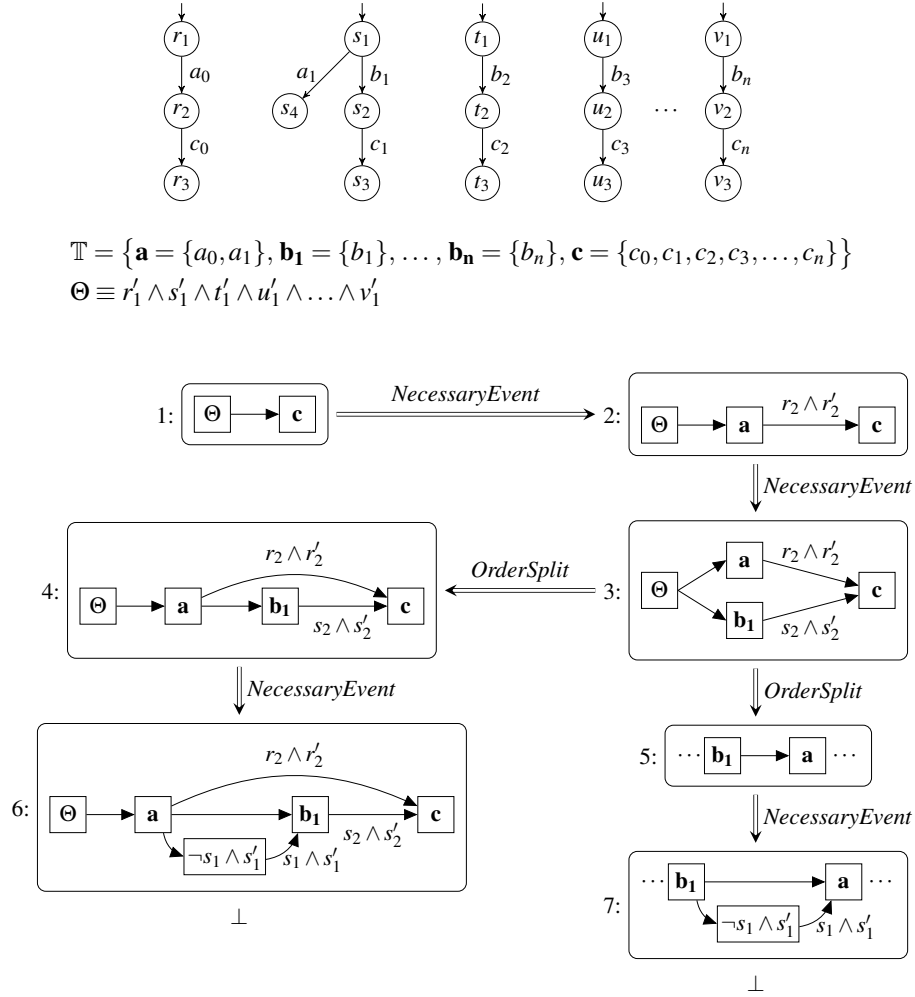

\vspace*{-3mm}
\paragraph{Safety.} Consider the synchronized system shown in the top part of Figure~\ref{fig:unfoldings}: the example was introduced by Esparza and Heljanko in \cite{Esparza/Unfoldings} to illustrate the exponential succinctness of Petri net unfoldings. There are $n+1$ processes, and we want to check whether the global transition $\bvar{c}$ is executable. Note that the state space of this system is exponential with respect to $n$: the system contains $3\cdot 2^{n-1}$ reachable states. Thus, approaches based on state space exploration will suffer from the state space explosion problem. The authors of \cite{Esparza/Unfoldings} show that the Petri net unfolding of the example system contains $2 \cdot n +3$ places, i.e., a \emph{linear size} unfolding can represent succinctly the exponential state space.

We use the same example to demonstrate that the trace unwinding of the example system never exceeds $n + 6$ nodes, but a \emph{constant size} unwinding of just $7$ nodes also suffice, which we show in the bottom part of Figure~\ref{fig:unfoldings}. Node~1, the root of the unwinding, captures all system traces where $\bvar{c}$ is executed. One of it's preconditions is that the first process should be at location $r_2$; but the initial condition says that the system is at location $r_1$: a contradiction. Thus, a transition that goes from $r_1$ to $r_2$ is \emph{necessary}, and we insert the only such transition, $\bvar{a}$, into the trace of node~2. Formally, this is done by applying the trace transformer \LastNecessaryEvent, this is where the link label $r_2 \land r_2'$ comes from: it enforces to select the last transition that goes into location $r_2$. By a similar reasoning we conclude that transition \bvar{b_1} is also necessary, and include it into the trace of node~3. Notice that these events occur concurrently, i.e. no specific order between them is specified till now.

In the next iteration, when we try to put them in some linear order, we find out that these transitions contradict each other: \bvar{a} goes to location $s_4$, \bvar{b_1} goes to location $s_2$, but both can start only at location $s_1$. Therefore, we perform a case split  between two possible linearizations using the transformer \OrderSplit, and obtain the traces of nodes 4 and 5. In node~4  the contradiction between now linearly ordered transitions \bvar{a} and \bvar{b_1} is enforced by the trace, and we again apply the transformer \LastNecessaryEvent. It inserts an event between transitions \bvar{a} and \bvar{b_1} which needs to change the location from $s_4$ (the postcondition of \bvar{a}) to $s_1$ (the precondition of \bvar{b_1}). Formally, this requirement is captured by the Craig interpolant between between the post- and pre-conditions of \bvar{a} and \bvar{b_1}, respectively, which happens to be $\neg s_1$. As there is no transition that goes from a location different from $s_1$ to $s_1$, this trace is declared as contradictory, and the left branch of the unwinding is closed. For the right branch of the unwinding, consisting of nodes 5 and 7, we proceed in the same way, and also close it as contradictory. Thus, the unwinding is complete, and we conclude that transition \bvar{c} is not executable.

The unwinding of Figure~\ref{fig:unfoldings} has constant size, independent on the number of processes in the example system. In the worst case it could have linear size: for that to happen, the other necessary transitions \bvar{b_2} through \bvar{b_n} would have to be introduced into concurrent before transitions \bvar{a}  and \bvar{b_1} are introduced. 
%But note that these transitions do not form any contradictions both with each other as well as with \bvar{a}  and \bvar{b_1}; therefore, they would stay concurrent in any trace that involves them, and no ordering splits would be necessary. 
In fact, a simple heuristic is able to select transitions \bvar{a}  and \bvar{b_1} first, and it will always produce the trace unwinding of constant size.

\vspace*{-3mm}
\paragraph{Liveness.}

\renewcommand{\t}{\hspace*{3mm}}

\begin{figure}
\centering
\scalebox{0.8} {
\begin{tabular}{p{4.5cm}  p{4.5cm} p{4.5cm} p{4.5cm}}
\textbf{Producer 1} & \textbf{Producer 2} & \textbf{Consumer 1} & \textbf{Consumer 2} \\

\parbox[t]{4.5cm} {
\texttt{while (p1>0) $\lbrace$
      \\\t if(*)  q1++;
      \\\t else \hspace*{2mm}q2++;
      \\\t p1--;
      \\$\rbrace$
}
}
 & 

\parbox[t]{4.5cm} {
\texttt{while (p2>0) $\lbrace$
      \\\t if(*)  q1++;
      \\\t else \hspace*{2mm}q2++;
      \\\t p2--;
      \\$\rbrace$
}
}

&

\parbox[t]{4.5cm} {
\texttt{while (true) $\lbrace$
      \\\t await(q1>0);
      \\\t skip; //step 1
      \\\t skip; //step 2
      \\\t q1--;
      \\$\rbrace$
}
}

&

\parbox[t]{4.5cm} {
\texttt{while (true) $\lbrace$
      \\\t await(q2>0);
      \\\t skip; //step 1
      \\\t skip; //step 2
      \\\t q2--;
      \\$\rbrace$
}
} 
 \\

\end{tabular}

}

\vspace*{0mm}

\scalebox{0.8} {
    \begin{tikzpicture}[every state/.style= {inner sep=0.7mm,minimum size=0pt}, node distance=2.5cm,>=triangle 45]

      \node[state]    (s1)  at  (1, 3) {$1$};
      \node (sstart) [circle,left=0.4cm of s1] {};
      \node[state, right of= s1] (s2) {$2$};
      \node[state, right of= s2] (s3) {$3$};
      \path
      (sstart)  edge [->]  (s1)
      (s1)  edge [->] node[above,yshift=1mm] {$\boldsymbol{a_1}: p_1>0$} (s2)
      (s2)  edge [->,bend left] node[above,yshift=1mm] {$\boldsymbol{a_2}: q_1'=q_1+1$} (s3)
      (s2)  edge [->,bend right] node[below,xshift=-7mm] {$\boldsymbol{a_3}: q_2'=q_2+1$} (s3)
      (s3)  edge [->,bend left=60]  node[below,yshift=-1mm] {$\boldsymbol{a_4}: p_1'=p_1-1$} (s1)
      ;

      \node[state, right=9cm of s1]    (s1)  at  (1, 3) {$1$};
      \node (sstart) [circle,left=0.4cm of s1] {};
      \node[state, right=2cm of s1] (s2) {$2$};
      \node[state, right=1.5cm of s2] (s3) {$3$};
      \node[state, right=1.5cm of s3] (s4) {$4$};
      \path
      (sstart)  edge [->]  (s1)
      (s1)  edge [->] node[above,yshift=1mm] {$\boldsymbol{c_1}: q_1>0$} (s2)
   %   (s2)  edge [->,bend left=45] node[below,yshift=-1mm] {$\boldsymbol{c_2}: q_1'=q_1-1$} (s1)
      (s2)  edge [->] node[above,yshift=1mm] {$\boldsymbol{c_2}: \mathit{true}$} (s3)
      (s3)  edge [->] node[above,yshift=1mm] {$\boldsymbol{c_3}: \mathit{true}$} (s4)
      (s4)  edge [->,bend left=45]  node[below,yshift=-1mm] {$\boldsymbol{c_4}: q_1'=q_1-1$} (s1)
      ;

       \end{tikzpicture}
}

\caption{The \emph{Producer-Consumer} benchmark, shown here for 2 producers and 2 consumers (\emph{Top}: pseudocode; \emph{Bottom}: control flow graphs with labeled transitions for Producer 1 and Consumer 1). The producer threads draw tasks from individual pools and distribute them to nondeterministically chosen queues, each served by a dedicated consumer thread; two steps are needed to process a task. The integer variables $p_1$ and $p_2$ model the number of tasks left in the pools of Producers~1 and 2, the integer variables $q_1$ and $q_2$ model the number of tasks in the queues of Consumers~1 and 2.}
\label{fig:prod-cons}
%\figureshrinker
\vspace*{-2mm}
\end{figure}
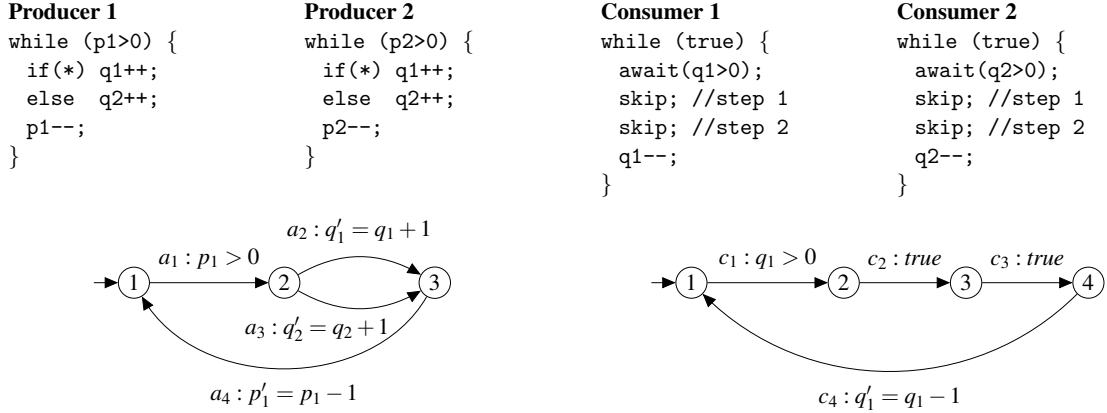

Consider the Producer-Consumer example presented in Figure~\ref{fig:prod-cons}, which  is a simplified model of the \emph{Map-Reduce} architecture from distributed processing: producers model the mapping step for separate data sources, consumers model the reducing step for different types of input data. The natural requirement for this architecture is that the processing terminates for any finite amount of input data.

\tikzstyle{box}=[align=center,circle,draw,inner sep=0.5mm, minimum height=6mm, minimum width=6mm]
\tikzstyle{rec}=[align=center,rectangle,draw,inner sep=1mm, minimum height=6mm]
\tikzstyle{lab}=[rectangle,draw,fill=white, above=-1mm]
\tikzstyle{leaf}=[circle,draw,thick,inner sep=1mm,minimum size=0mm]
\tikzstyle{arr}=[->,double,double distance=0.5mm,>=stealth new, arrow head=3mm]

\begin{figure}[!t]
\centering
\hspace*{-10mm}
\scalebox{0.8} {
\begin{tikzpicture}[>=triangle 45, node distance=10mm]

  \node(trace2) {
  \begin{tikzpicture}[node distance=8mm]
  %\node[box] (I) {$I$};
  \node (I) {\large $1:$};
  \node[right=0mm of I,inner sep=0mm](lb) {\Huge $($};
  \node[box, right=1mm of lb] (i) {$\top$};
  \node[inner sep=0mm,right=1mm of i] (rb) {\Huge $)$};
  \node[inner sep=0mm,right=0mm of rb,yshift=2mm] (om) {\huge $^\omega$};
  \path
  %(I)  edge [->]  (lb)
  ;
  \end{tikzpicture}
  };

  \node[below right=of trace2](trace3) {
  \begin{tikzpicture}[node distance=8mm]
 % \node[box] (I) {$I$};
  \node (I) {\large $5:$};
  \node[right=0mm of I,inner sep=0mm](lb) {\Huge $($};
  \node[box, right=1mm of lb] (i) {$c_1$};
  \node[inner sep=0mm,right=1mm of i] (rb) {\Huge $)$};
  \node[inner sep=0mm,right=0mm of rb,yshift=2mm] (om) {\huge $^\omega$};
  \path
  %(I)  edge [->]  (lb)
  ;
  \end{tikzpicture}
  };

  \node[below=of trace3](trace4) {
  \begin{tikzpicture}[node distance=8mm]
  %\node[box] (I) {$I$};
  \node (I) {\large $6:$};
  \node[right=0mm of I,inner sep=0mm](lb) {\Huge $($};
  \node[box, right=0mm of lb] (c1) {$c_1$};
  \node[box, right=of c1] (c2) {$c_2$};
  \node[box, right=of c2] (c3) {$c_3$};
  \node[box, right=of c3] (c4) {$c_4$};
  \node[inner sep=0mm,right=0mm of c4] (rb) {\Huge $)$};
  \node[inner sep=0mm,right=0mm of rb,yshift=2mm] (om) {\huge $^\omega$};
  \path
  %(I)  edge [->]  (lb)
  (c1)  edge [->]  (c2)
  (c2)  edge [->]  (c3)
  (c3)  edge [->]  (c4)
  ;

  \node[below=3mm of c2,xshift=6mm] {\large $\mathit{Terminating}: q_1$};

  \end{tikzpicture}
  };

  \node[below=of trace4](trace45) {
  \begin{tikzpicture}[node distance=8mm]
  %\node[box] (I) {$I$};
  \node (I) {\large $7:$};
  \node[right=0mm of I,inner sep=0mm,scale=1.8](lb) {\Huge $($};
  \node[box, right=0mm of lb,yshift=5mm] (c1) {$c_1$};
  \node[box, right=of c1] (c2) {$c_2$};
  \node[box, right=of c2] (c3) {$c_3$};
  \node[box, right=of c3] (c4) {$c_4$};
  \node[rec, below=3mm of c2,xshift=4mm] (a2) {$q_1'>q_1$};
  \node[inner sep=0mm,right=48mm of lb,scale=1.8] (rb) {\Huge $)$};
  \node[inner sep=0mm,right=0mm of rb,yshift=6mm,xshift=-2mm] (om) {\huge $^\omega$};
  \path
  %(I)  edge [->]  (lb)
  (c1)  edge [->]  (c2)
  (c2)  edge [->]  (c3)
  (c3)  edge [->]  (c4)
  ;
  \end{tikzpicture}
  };

  \node[below=of trace45](trace5) {
  \begin{tikzpicture}[node distance=8mm]
  %\node[box] (I) {$I$};
  \node (I) {\large $8:$};
  \node[right=0mm of I,inner sep=0mm,scale=1.8](lb) {\Huge $($};
  \node[box, right=0mm of lb,yshift=5mm] (c1) {$c_1$};
  \node[box, right=of c1] (c2) {$c_2$};
  \node[box, right=of c2] (c3) {$c_3$};
  \node[box, right=of c3] (c4) {$c_4$};
  \node[box, below=3mm of c2,xshift=4mm] (a2) {$a_2$};
  \node[inner sep=0mm,right=48mm of lb,scale=1.8] (rb) {\Huge $)$};
  \node[inner sep=0mm,right=0mm of rb,yshift=6mm,xshift=-2mm] (om) {\huge $^\omega$};
  \path
  %(I)  edge [->]  (lb)
  (c1)  edge [->]  (c2)
  (c2)  edge [->]  (c3)
  (c3)  edge [->]  (c4)
  ;
%  \node[right=5mm of rb] {};
  \end{tikzpicture}
  };

  \node[below=of trace5](trace6) {
  \begin{tikzpicture}[node distance=8mm]
  %\node[box] (I) {$I$};
  \node (I) {\large $9:$};
  \node[right=0mm of I,inner sep=0mm,scale=1.8](lb) {\Huge $($};
  \node[box, right=0mm of lb,yshift=5mm] (c1) {$c_1$};
  \node[box, right=of c1] (c2) {$c_2$};
  \node[box, right=of c2] (c3) {$c_3$};
  \node[box, right=of c3] (c4) {$c_4$};
  \node[box, below=3mm of c1,xshift=6mm] (a2) {$a_1$};
  \node[box, right=of a2] (a4) {$a_2$};
  \node[box, right=of a4] (a1) {$a_4$};
  \node[inner sep=0mm,right=48mm of lb,scale=1.8] (rb) {\Huge $)$};
  \node[inner sep=0mm,right=0mm of rb,yshift=6mm,xshift=-2mm] (om) {\huge $^\omega$};
  \path
  %(I)  edge [->]  (lb)
  (c1)  edge [->]  (c2)
 (c2)  edge [->]  (c3)
  (c3)  edge [->]  (c4)
  (a2)  edge [->]  (a4)
  (a4)  edge [->]  (a1)
  ;
  \end{tikzpicture}
  };

  \node[below left=of trace2,xshift=-5mm](trace7) {
  \begin{tikzpicture}[node distance=8mm]
  %\node[box] (I) {$I$};
  \node (I) {\large $2:$};
  \node[right=0mm of I,inner sep=0mm,](lb) {\Huge $($};
  \node[box, right=1mm of lb] (i) {$a_1$};
  \node[inner sep=0mm,right=1mm of i] (rb) {\Huge $)$};
  \node[inner sep=0mm,right=0mm of rb,yshift=2mm] (om) {\huge $^\omega$};
  \path
  %(I)  edge [->]  (lb)
  ;
  \end{tikzpicture}
  };

  \node[below=of trace7](trace8) {
  \begin{tikzpicture}[node distance=8mm]
  %\node[box] (I) {$I$};
  \node (I) {\large $3:$};
  \node[right=0mm of I,inner sep=0mm](lb) {\Huge $($};
  \node[box, right=1mm of lb] (a1) {$a_1$};
  \node[box, right=of a1] (a4) {$a_4$};
  \node[inner sep=0mm,right=1mm of a4] (rb) {\Huge $)$};
  \node[inner sep=0mm,right=0mm of rb,yshift=2mm] (om) {\huge $^\omega$};
  \path
  %(I)  edge [->]  (lb)
  (a1)  edge [->]  (a4)
  ;
  \node[below=3mm of a1,xshift=7mm] {\large $\mathit{Terminating}: p_1$};
  
  \end{tikzpicture}
  };

  \node[below=of trace8](trace9) {
  \begin{tikzpicture}[node distance=8mm]
  %\node[box] (I) {$I$};
  \node (I) {\large $4:$};
  \node[right=0mm of I,inner sep=0mm,scale=1.8](lb) {\Huge $($};
  \node[box, right=0mm of lb,yshift=5mm] (a1) {$a_1$};
  \node[box, right=of a1] (a4) {$a_4$};
  \node[rec, below=3mm of a1,xshift=6mm] {$p_1'>p_1$};
  \node[inner sep=0mm,right=20mm of lb,scale=1.8] (rb) {\Huge $)$};
  \node[inner sep=0mm,right=0mm of rb,yshift=6mm,xshift=-2mm] (om) {\huge $^\omega$};
  \path
  %(I)  edge [->]  (lb)
  (a1)  edge [->]  (a4)
  ;

  \node[below=13mm of a1,xshift=6mm] {\large $\bot$};

  \end{tikzpicture}
  };

  \node[left= 50mm of trace6,yshift=5mm](dependencies) {
  \begin{tikzpicture}
  %\node[box] (I) {$I$};
  \node (p1) {\large $p_1$};
  \node[right=10mm of p1] (p2) {\large $p_2$};
  \node[below=15mm of p1] (q1) {\large $q_1$};
  \node[below=15mm of p2] (q2) {\large $q_2$};

  \path
  (q1)  edge [->]  (p1)
  (q1)  edge [->]  (p2)
  (q2)  edge [->]  (p1)
  (q2)  edge [->]  (p2)
  ;

  \end{tikzpicture}
  };

  \node[below=of trace2](dots1) {\large$\ldots$};
  \node[below=5mm of trace45,xshift=-10mm](dots2) {\large$\ldots$};

  \path
  (trace2)  edge [arr] node[right,yshift=3mm] {\large\textit{Instantiate$^\omega$}} (trace3)
  (trace2)  edge [arr]  (dots1)
  (trace3)  edge [arr] node[right,xshift=2mm] {\large\textit{NecessaryCycleEvent}} (trace4)
  (trace4)  edge [arr] node[right,xshift=2mm] {\large\textit{InvarianceSplit}} (trace45)
  (trace45)  edge [arr] node[right,xshift=2mm] {\large\textit{Instantiate$^\omega$}} (trace5)
  (trace45)  edge [arr]  (dots2)
  (trace5)  edge [arr] node[right,xshift=2mm] {\large\textit{NecessaryCycleEvent}} (trace6)
  (trace2)  edge [arr] node[left,yshift=3mm] {\large\textit{Instantiate$^\omega$}} (trace7)
  (trace7)  edge [arr] node[left,xshift=-2mm] {\large\textit{NecessaryCycleEvent}} (trace8)
  (trace8)  edge [arr] node[left,xshift=-2mm] {\large\textit{InvarianceSplit}} (trace9)
  %(trace6.west)  edge [arr, bend left=80,dashed]  (trace7)
  ;
  \draw[arr,dashed] (trace6.west) .. controls (trace9) and (trace4) .. (trace7.east);
    
\end{tikzpicture}
}

\caption{Termination proof for the Producer-Consumer example of Figure~\ref{fig:prod-cons}. \emph{Bottom left}: partially ordered ranking function discovered in the analysis.}
\label{fig:prod-cons-analysis}
\vspace*{-3mm}
\end{figure}
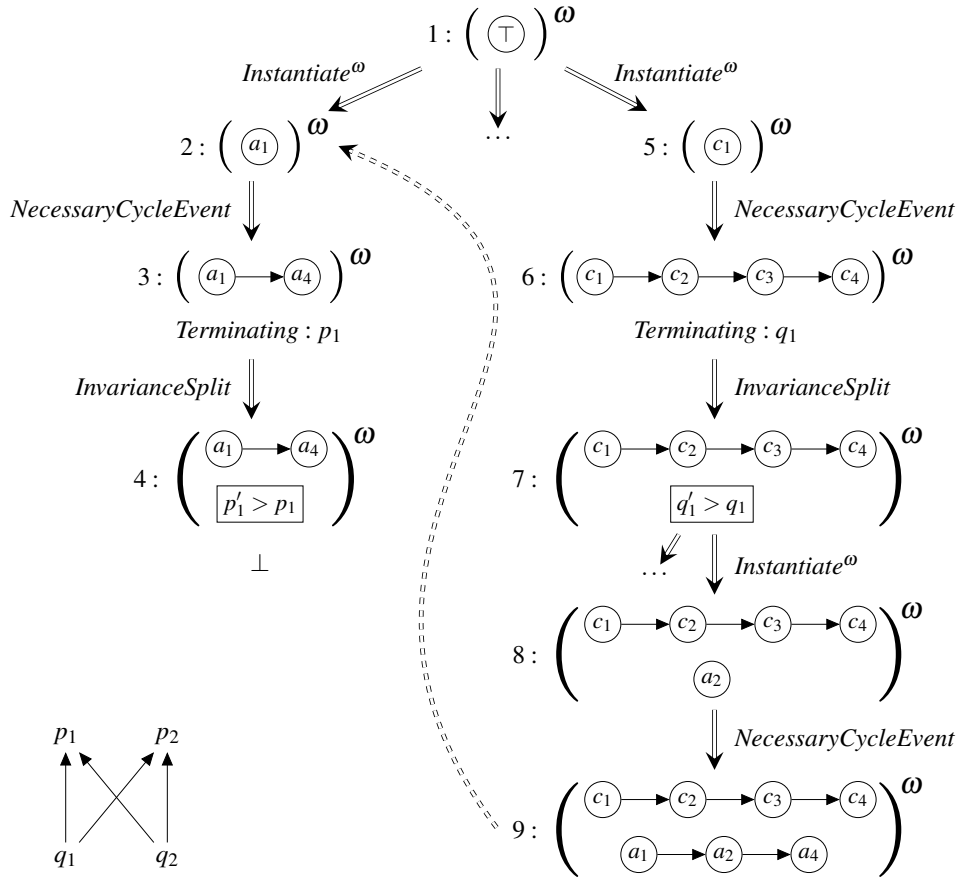

Our analysis starts with the assumption (by way of contradiction) that there exists some infinite computation. The assumption is expressed as the concurrent trace of node 1 in Figure~\ref{fig:prod-cons-analysis}: infinitely often some transition should occur. The transition is so far unknown, and therefore characterized by the predicate $\top$. 
Our argument proceeds by instantiating this unknown transition with the transitions of the program,
resulting in one new trace per transition. %The $\Instantiate^\omega$ trace transformer represents a case distinction, and we will need to discharge all cases.

For example, transition $a_1$ of Producer 1, gives us the trace of node 2. 
The consequence of the decision that $a_1$ occurs infinitely often is that $a_4$ must also occur infinitely often: 
%after the execution of $a_1$, the program counter of producer 1 equals 2, and the precondition for the execution of $a_1$ is that it is equal to 1. 
$a_1$ starts at location 1, and only $a_4$ can bring Producer~1 to this location. 
%The only transition that can achieve that goal is $a_4$.
% (here we oversimplify to make the presentation clearer; in the algorithm we derive the necessity of event $a_4$ by the interpolation-based local safety analysis). 
The requirement that both $a_1$ and $a_4$ occur infinitely often is expressed as the trace of node 3, obtained from the trace of node 1 by the \NecessaryCycleEvent\ trace transformer. %The edge between $a_1$ and $a_4$ specifies an ordering between the two transitions; between them, there may be an arbitrary number of other transitions.
The trace of node 3 is terminating: $p_1$ is decreased infinitely often and is bounded from below; it is therefore a ranking function. An infinite computation might exist only if some transition increases $p_1$ (satisfies the predicate $p_1'>p_1$) is executed infinitely often. This situation is expressed by the trace of node 4, obtained by the application of the \InvarianceSplit\ trace transformer.
Since there is no transition in the program transition relation that satisfies $p_1'>p_1$, we arrive at a contradiction.

% Notice how in the above step we transformed one liveness condition ($a_1$ and $a_4$ should occur infinitely often) into another liveness condition (event satisfying $p_1'>p_1$ should occur infinitely often): we stay in the liveness territory as long as possible, before reverting to the safety reasoning, thus avoiding the bottleneck from which termination provers suffer the most.

Let us explore another instantiation of the unknown event in the trace of node 1, this time with transition $c_1$ of Consumer 1: we obtain the trace of node 5. Again, exploring causal consequences, local safety analysis gives us that events $c_2$, $c_3$, and $c_4$ should also occur infinitely often in the trace: we insert them, and get the trace of node 6. Termination analysis for that trace gives us the ranking function $q_1$: it is bounded from below by event $c_1$ and decreased by event $c_4$. Again, we conclude that the event increasing $q_1$ should occur infinitely often, and introduce it in the trace of node 7. 

Next, we try all possible instantiations of the event characterized by the predicate $(q_1'>q_1)$: there are two transitions that satisfy the predicate, namely $a_2$ and $b_2$. We explore the instantiation with $a_2$ in the trace of node 8,
%. The local safety analysis allows us to conclude that, 
and see that transitions $a_1$ and $a_4$ should occur infinitely often (node 9). At this point, we realize that the trace of node 9 contains as a subgraph the trace of node 2, namely the transition $a_1$.
% We can conclude, without repeating the analysis done of nodes 2--4, that there is no infinite computation corresponding to the trace of node~9.
We cover node~9 with node 2, and avoid repeating the analysis done for nodes 2--4.
The remaining tableau branches are analyzed similarly. The resulting tableau for the case of two producers and two consumers will have the shape shown in the bottom left part of Figure~\ref{fig:prod-cons-analysis}.
It can be interpreted as a partially ordered ranking function, which shows that all threads satisfying the function components $p_1$ and $p_2$ terminate unconditionally, while the threads that satisfy the function components $q_1$ and $q_2$ terminate under the condition that both of the previous components terminate. 
Notice also that the tableau is of quadratic size with respect to the number of threads.

\section{Conclusions}

Causality-based model checking has significant advantages over
standard state-based model checking. As illustrated by the 
two examples, the complexity of the verification problem can be
substantially lower; in particular, the complexity of verifying
multi-threaded programs with locks
reduces from exponential to polynomial (see \cite{Safety}, Theorem 4).  The efficiency of causality-based
model checking is also reflected in the experimental results obtained with our
tool implementation \textsc{Arctor} (cf.~\cite{Liveness}). In our experience,
\textsc{Arctor} scales to many, even hundreds, of parallel threads in
benchmarks where other tools can only handle a small number of
parallel threads or no parallelism at all: see Table~\ref{table} for
experimental results obtained with \textsc{Arctor} on the
Producer-Consumer example.

\begin{table}[t]
  \centering
\hspace*{-1mm}
\scalebox{0.84}{
  \begin{tabular}{| c || c | c || c | c || c | c || c | c | c | }
    \hline
     & \multicolumn{2}{|c||}{Terminator}  & \multicolumn{2}{c||}{T2} & \multicolumn{2}{c||}{AProVE} & \multicolumn{3}{c|}{\textsc{Arctor}}  \\ \hline
    Threads  & Time(s) & Mem.(MB) & Time(s) & Mem.(MB) & Time(s) & Mem.(MB) & Time(s) & Mem.(MB) & Vertices \\ \hline
    \hline
     1        &  3.37   &  26      &  2.42    &  38     &  3.17   & 237     & 0.002   &   2.3    &  6      \\ %\hline
     2        &  1397   &  1394    &  3.25    &  44     &  6.79   & 523     & 0.002   &   2.6    &  11     \\ %\hline
     3        &$\times$ &  MO      &  U(29.2) &  253    & U(26.6) & 1439    & 0.002   &   2.6    &  21     \\ %\hline
     4        &$\times$ &  MO      &  U(36.6) &  316    & U(71.2) & 1455    & 0.003   &   2.7    &  30     \\ %\hline
     5        &$\times$ &  MO      &  U(30.7) & 400     & U(312)  & 1536    & 0.007   &   2.7    &  44     \\ %\hline
%    10        &$\times$ &  MO      &  Z3-TO   &$\times$ &$\times$ & MO      & 0.027   &  3.0     &  135    \\ %\hline
    20        &$\times$ &  MO      &  Z3-TO   &$\times$ &$\times$ & MO      & 0.30    &  4.2     &  470    \\ %\hline
    40        &$\times$ &  MO      &  Z3-TO   &$\times$ &$\times$ & MO      & 4.30    & 12.7     & 1740    \\ %\hline
    60        &$\times$ &  MO      &  Z3-TO   &$\times$ &$\times$ & MO      & 20.8    & 35       & 3810    \\ %\hline
    80        &$\times$ &  MO      &  Z3-TO   &$\times$ &$\times$ & MO      & 67.7    & 145      & 6680    \\ %\hline
   100        &$\times$ &  MO      &  Z3-TO   &$\times$ &$\times$ & MO      & 172     & 231      & 10350   \\ %\hline
  \hline
  \end{tabular}
}
  \vspace*{2mm}
  \caption{Running times of the termination provers \emph{Terminator}, \emph{T2}, \emph{AProVE}, and \textsc{Arctor} on the Producer-Consumer benchmark~\cite{Liveness}. MO stands for memout. U indicates that the termination prover returned ``unknown''; Z3-TO indicates a timeout in the Z3 SMT solver.}
  \label{table}
\end{table}

%\nocite{*}
\bibliographystyle{eptcs} \bibliography{references}

\end{document}